# AWAPart: Adaptive Workload-Aware Partitioning Knowledge Graphs

Amitabh Priyadarshi
Department of Computer Science
University of Georgia
Athens, GA, USA
email: amitabh.priyadarshi@uga.edu

Krzysztof J. Kochut
Department of Computer Science
University of Georgia
Athens, GA, USA
email: kkochut@uga.edu

*Abstract*—Large-scale knowledge graphs are increasingly common in many domains. Their large sizes often exceed the limits of systems storing the graphs in a centralized data store, especially if placed in main memory. To overcome this, large knowledge graphs need to be partitioned into multiple sub-graphs and placed in nodes in a distributed system. But querying these fragmented sub-graphs poses new challenges, such as increased communication costs, due to distributed joins involving cut edges. To combat these problems, a good partitioning should reduce the edge cuts while considering a given query workload. However, a partitioned graph needs to be continually re-partitioned to accommodate changes in the query workload and maintain a good average processing time. In this paper, an adaptive partitioning method for large-scale knowledge graphs is introduced, which adapts the partitioning in response to changes in the query workload. Our evaluation demonstrates that the performance of processing time for queries is improved after dynamically adapting the partitioning of knowledge graph triples.

## I. INTRODUCTION

The availability of large-scale knowledge graphs, which often holds hundreds of millions of vertices and edges, such as the ones used in social network systems or in other real-world systems, requires large-scale graph processing. Of-ten, these datasets are too large to be stored and processed in a centralized data store, especially if it is maintained in main memory. Instead, the knowledge graph often needs to be partitioned into multiple sub-graphs, called shards, and transferred to multiple nodes in a distributed system. However, these systems frequently suffer from network latency. One of the techniques to improve the query answering performance is to reduce the inter-process communication between graph processing subsystems. While graph partitioning may be an effective pre-processing technique to improve the runtime performance, the cost of frequent partitioning of the entire large-scale knowledge graph may be prohibitive. In this case, it would be advantageous to partition the graph only once, initially, and make only necessary partitioning adjustments, afterwards. For example, such adjustments are needed in case of changes to the query workload.

First, let us talk about graph partitioning. Given a graph $G = (V, E)$, where $V$ is a set of vertices and $E$ is a set of edges and a number $k > 1$, a *graph partitioning of G* is a subdivision of vertices of $G$ into *subsets* of vertices $V_1, ..., V_k$ that partition the set V. A *balance constraint* requires that all partition blocks are equal, or close, in size. In addition, a common objective function is to minimize the *total number of cuts*, i.e., edges crossing (cutting) partition boundaries.

Our knowledge graph dataset is in the form of Resource Description Framework (RDF) [1]. RDF enables the embedding of machine-readable information on the web. A resource can be represented using a URL on the web. The RDF statement which comprises of three parts called a triple, consists of (s, p, o) resource, property, and value of resource. RDF Schema [2] defines Classes and Properties that create a taxonomy for arranging the RDF data. Web Ontology Language (OWL) [3] is a language to describe complex knowledge about the things and provides a way to represent the relationships between a group of things. The documents in OWL are known as Ontologies [4]. The RDF query language SPARQL is the W3C standard that is used for querying the data in RDF graphs for exploring relationships between resources. SPARQL tries to match a triple pattern in an RDF graph. A SPARQL endpoint accepts SPARQL queries that return the result via HTTP. The partitioned RDF graph can be accessed through different SPARQL endpoints in a single query using Federated SPARQL Query [5]. The SERVICE keyword is used to direct a portion of a query towards a particular SPARQL endpoint. In Table 1 there is an example of LUBM's [22] SPARQL query and its federated query. The federated query processor merges the results coming from the various SPARQL endpoints.

TABLE I. ORIGINAL AND FEDERATED QUERY OF LUBM 9TH QUERY

| Original Query | Federated Query |
|---|---|
| SELECT ?X ?Y ?Z FROM lubm WHERE{ <br> ?X rdf:type ub:Student. <br> ?Y rdf:type ub:Faculty . <br> ?Z rdf:type ub:Course . <br> ?X ub:advisor ?Y . <br> ?Y ub:teacherOf ?Y . <br> ?X ub:takesCourse ?Z . <br> } | SELECT ?X ?Y ?Z FROM lubm WHERE { <br> ?X rdf:type ub:Student. <br>   SERVICE \<Sparql endpoint\> {?Y rdf:type ub:Faculty .} <br> ?Z rdf:type ub:Course . <br>   SERVICE \<Sparql endpoint\> {?X ub:advisor ?Y . <br>   SERVICE \<Sparql endpoint\> {?Y ub:teacherOf ?Y .} <br> ?X ub:takesCourse ?Z . <br> } |

This paper is outlined as follows. Section 2 provides an overview of related work. Section 3 discusses the partitioning method. Section 4 is about the architecture and workflow of the system. Section 5 is dedicated for the experiments, and Section 6 concludes the paper.

## II. RELATED WORK

Usually, partitioning a large-scale graph decreases query processing efficiency. However, this decrease can be mitigated if the partitioning is adjusted to a query workload and tuned to reduce the workload demands for inter-partition communication. Related work on graph partitioning and its implication on query processing is addressed in this section.

The graph partitioning problem is NP-complete [6]. Many practical techniques have been developed to address this issue, including spectral partitioning methods [7] and geometric partitioning methods [8]. Barnard and Simon [9] proposed the multilevel method to graph partitioning, and Hendrickson and Leland [10] enhanced it. Coarsening, initial partitioning, and uncoarsening are the three basic phases of the multilevel technique. Karypis et al. [11] employ a recursive multilevel bisection method for graph bisection to generate a k-partition on the coarsest level in their partitioning approach.

Workload-aware, distributed RDF systems include DREAM [12], WARP [13], PARTOUT [14], AdPart [15], and WISE [16]. DREAM [12] only partitions SPARQL queries into subgraph patterns, not the entire RDF dataset. The RDF dataset is replicated among nodes. It is designed in a master-slave architecture, with each node using RDF-3X [17] on its assigned data for statistical estimation and query evaluation. WARP [13] assigns each vertex of the RDF graph to a partition using the underlying METIS system. The triples are subsequently assigned to partitions, which are then stored in a triple store on dedicated hosts (RDF-3X). WARP uses an n-hop distance to compute the query's center node and radius. If the query is within n-hops, WARP sends the query to all partitions to be executed in parallel. A complex question is broken down into multiple sub-queries, which are then run in parallel, and the results are merged. PARTOUT [14] uses normalization and anonymization to extract representative triple patterns from a query workload by substituting infrequent URIs and literals with variables. Frequent URIs (above a frequency threshold) are normalized. PARTOUT uses an adapted version of RDF-3X as a triple store for their n hosts. AdPart [15] is an in-memory RDF system that incrementally re-partitions RDF data. In an in-memory data structure, each worker stores its local set of triples. AdPart provides an ability to monitor and index the workloads in the form of hierarchical heat maps. It introduces Incremental ReDistribution (IRD), which is a query workload-guided combination of hash partitioning and k-hop replication. WISE [16] is a workload-aware, runtime-adaptive partitioning system for large-scale knowledge graphs. Based on changes in the workload, a partitioning can be modified incrementally by trading triples. The frequencies of SPARQL queries are kept in a Query Span structure. When migrating the triples, a cost model that maximizes the migration gain while preserving the balanced partition is applied.

AWAPart, presented in this paper, is a query-adaptive workload-aware knowledge graph partitioning algorithm that extracts features from both the query workload and the dataset. These features are utilized to create a distance matrix between queries and then cluster similar queries together using hierarchical agglomerative clustering. From the knowledge graph data, subgraphs (partitions) associated with these features are produced and distributed as shards in a computing cluster. The partitioning of the graph will be adjusted in response to changes in the workload, e.g., if some queries are replaced or their execution frequencies change. This is done by updating metadata with new features from new queries. These new features are being clustered again. Scoring helps the system to swap data associated with features from one shard to another. This swapping is done to reduce the edge cuts and minimize query runtime. Importantly, unlike the related systems, ours does not rely on a specialized data store implementation and uses an *off-the-shelf* knowledge graph storage and query processing system (Virtuoso [18]) and relies on standard SPARQL queries for distributed processing.

## III. Workload-Aware Adaptive Knowledge Graph Partitioning

As the workload changes over time, an optimized (current) graph partition eventually becomes inefficient for the modified workload. AWAPart's goal is to adapt an existing knowledge graph partitioning to changes in the query workload, to optimize the workload processing time. Critical features from the current and modified workloads are extracted and analyzed. Features of queries in the changed workload are clustered, based on the similarity measures. The features in the new and old clusters are compared and a new optimized partition is created. The system then dynamically adjusts the deployed partitioning (shards) by exchanging triples belonging to the modified features between shards in the cluster. However, the analysis of the workload and the resulting adjustment of the partitions (shards) is infrequent. It can be performed in the background, without interrupting the process of querying. Queries in the workload are re-written to form federated SPARQL queries for processing on the cluster. Adjusting the partitioning (shards) aims to limit the number of distributed joins (utilizing triples from different shards), which decreases workload processing time.

### A. Query Feature Extraction

The query feature metadata maintains the information about the triple patterns, which is referred as features in this paper, present in a set of triples. This metadata is maintained for each shard to describe the current set of triples in the shard.

The following features are used to describe various triple patterns, which are identified for the purpose of query workload clustering.
- Property (P): This feature represents all triples which share a given predicate P (triple's property).
- Property-Object (PO): This feature represents all triples sharing the same predicate P and the object (triple's property and object).

Other feature types used for query analysis are:
- Subject-Subject Join (SSJ): Triples sharing the same subject.
- Object-Object Join (OOJ): Triples sharing the same object.
- Object-Subject Join (OSJ): Triples connected on an entity which is the object in one triple and the subject in the other (it is referred as an "elbow" join in this paper).

We created the QueryAnalyzer which extracts the above features from the queries and creates the feature metadata.

This metadata represents the features, their frequencies, neighboring features, related data sizes and distributed joins in that query. This helps the system to optimize the partition by re-adjusting the partitioning based on the updated features. Currently, our QueryAnalyzer is built for the SPARQL query language, but it can be easily adapted to a different graph pattern-based query language, such as Cypher, used in the Neo4j [19] graph database.

Triples in the entire knowledge graph are indexed based on their subject, predicate and object, and the graph can be searched using any of them. For instance, it is easy to materialize the predicate feature and locate all triples with a given property P or any other triple pattern using a feature discussed above. For indexing the initial dataset of N-Triples, Apache Lucene API [20] is used to accelerate searching for triple features, while creating an initial partition [21] tailored to the initial query workload.

### B. Query Workload Clustering and Knowledge Graph Adaptive Partitioning

The distance matrix is used as an input data for data mining, such as multi-dimensional scaling, hierarchical clustering, etc. To measure the similarity between queries in a workload, based on their features, Jaccard similarity is used which generates a distance matrix. Clustering uses this distance matrix. The Jaccard similarity of sets A and B is the ratio of the intersection of sets A and B to the union of sets A and B. $J_{SIM} = |A \cap B| / |A \cup B|$.

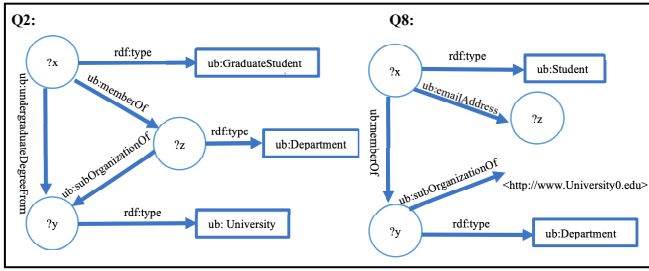

Figure 1. Distance between Q2 and Q8 is $1-J_{sim} = 1- (|Q2 \cap Q8|/|Q2 \cup Q8|)$
$= (1-3/8) = 0.625$

In Figure 1, query 2 has 6 features: (3 *PO* features: *rdf:type→ ub:GraduateStudent, rdf:type→ ub:Department, rdf:type→ ub:University* and 3 *P* features: *ub:memberOf, ub:subOrganizationOf, ub:underGraduateDegreeFrom*) while query 8 has 5 features (2 *PO* features: *rdf:type→ub:Student, rdf:type→ ub:Department,* and 3 *P* features: *ub:emailAddress, ub:subOrganizationOf, ub:memberOf*). The Jaccard similarity, which is the ratio of the intersection of both sets to the union of both sets, is 3/8. Now, the distance between two similar sets should be 0 and the Jaccard similarity of two identical sets returns 1. Therefore, the distance between queries Q2 and Q8 is $(1-J_{SIM}(Q2, Q8)) = 1 - 3/8 = 0.625$.

We used the Hierarchical agglomerative clustering (HAC) algorithm (Figure 4), which is a method of creating a hierarchy of clusters in a bottom-up fashion. The creation of clusters is based on the measure of similarity between clusters and the selection of linkage method. The shortest pairwise distance between queries determines the grouping. The distance matrix is recalculated once the two most similar clusters are being grouped together. Jaccard is used to create this distance matrix. This distance matrix is used to start the HAC. Recalculation of the distance matrix is based on the choice of linkage from single, complete, or average (Figure 2). Single linkage is the proximity between two nearest neighbors, complete linkage is the proximity between the farthest neighbor and average linkage is arithmetic mean of all proximities between each object on each cluster with every object on another cluster. Running HAC using a single linkage on LUBM queries gives a dendrogram (Figure 3). Clustering is computed periodically, based on the changes in the query workload and generates new dendrograms.

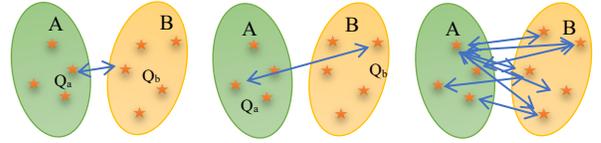

a. Single Linkage (SL)  b. Complete Linkage (CL)  c. Average Linkage (AL)

Figure 2.  a) $SL(A,B) = \min(D(Q_a,Q_b))$,  b) $CL(A,B) = \max(D(Q_a,Q_b))$ and c) $AL(A, B) = \frac{1}{n_A n_B}\sum_{a=1}^{n_A} \sum_{b=1}^{n_B} D(Q_a, Q_b)$

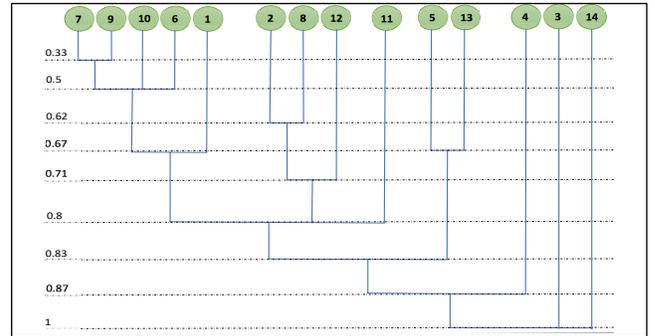

Figure 3.   HAC Dendrogram of LUBM's 14 Queries

| Input | Feature Distance Matrix ***D*** of workload Query |
|---|---|
| Output | HAC Dendrogram I |
| 1 | Assign for each D[n][n] into C[m] where m = n*n |
| 2 | **while** C.size > 1 **do** |
| 3 | **for** i = 1 **to** C.size **do** |
| 4 | **if** $(c_a,c_b) = min\ d(c_a,c_b)$ in C  //Distance *funct. d($c_1,c_2$)* |
| 5 | delete $c_a$ and $c_b$ from C |
| 6 | add $min\ d(c_a,c_b)$ in C |
| 7 | assign **I** = (old,  $c_a c_b,,\ min\ d(c_a,c_b)$)) |
| 8 | recalculate proximity matrix using (SL/CL/AL) **P = modifyDistance(** $c_a$ ,$c_b$ ,$min\ d(c_a,c_b)$**)** |
| 9 | for each P , $c_m$ = P[i][j] |
| 10 | Update C = $c_1, c_2, …, c_m$ |
| 11 | Output **I** |

Figure 4.   Hierarchical Agglomerative Clustering of Queries

The adaptive partitioning algorithm (Figure 5) takes the initial partitioning and a new query workload as input and outputs the partition minimizing the distributed joins, based on the new workload, with its new sets of features. To eliminate replication of data, only one copy of query features is stored in the shards. For removal of the replication and to decide in which shard the only copy of triples associated with the selected features will be transferred, the algorithm compares the statistics for each P or PO feature in each shard.

| Input | Initial Partition **P**, features **F$_G$**, New Queries workload **Q$_{new}$** |
|---|---|
| Output | Adaptive Partition **A** |
| 1 | Add queries **Q$_{new}$** and its frequency $f$ in Q$_{old}$ |
| 2 | Avg query execution time(T$_{base}$) = $(\sum_{Q=1}^{n}(\frac{\sum_{i=1}^{f} T_{Qi}}{f}))/n$ |
| 3 | Analyze Query **Q$_{new}$** for features **F$_{Qnew}$** |
| 4 | Run **HAC** on **F$_Q$**, where **F$_Q$** = **F$_{Qold}$** + **F$_{Qnew}$** |
| 5 | Create Feature set **g** based on HAC at similarity distance **d** |
| 6 | **Statistics (g, F$_Q$)** |
| 7 | Find key features **F$_K$** in **g**. |
| 8 | Find distributed joins of workload **D$_{Q(old+new)}$** = (D$_Q$ * $f$) |
| 9 | Find stats **S$_K$** for each **F$_K$** |
| 10 | Find p, q, s for shard C$_i$ and complete dataset T |
| 11 | **S$_K$** = $(p_c w_1 + q_c w_2 + s_c w_3) + (p_t w_4 + q_t w_5 + s_t w_6)$ //key features in p (peer features), in q (query), s (triple size) and $w_1$ to $w_6$ are weights. c and t are cluster and total. |
| 12 | Score for each **F$_K$** = $[min (D_{QR})*w * f] + S_K$  //D$_{QR}$(distributed joins of **F$_K$** in all query in every shard), $w$ (weight) and S$_K$ (key feature stat score). |
| 13 | **Balance_Partition (Score, g, F$_G$)** |
| 14 | select all F$_K$ from g with highest scores for each F$_K$ |
| 15 | Assign data associated to features set **g** into **P'**. |
| 16 | **Proximity_Query ()** |
| 17 | Find **F$_{prox}$** = proximity of **F$_{Unclustered}$** With **F$_{Clustered}$** |
| 18 | Assign max(**F$_{prox}$**) in cluster P$_i$' with their neighbor F. |
| 19 | Assign F$_X$ = F$_X$ + remaining F$_U$ |
| 20 | while F$_X$ not empty do |
| 21 | **P'$_{min}$** = Find min(**P'**) by size of data |
| 22 | **F$_{max}$** = Find max(**F** in **F$_X$**) by size of data |
| 23 | -Assign **F$_{max}$** into **P'$_{min}$** |
| 24 | Avg execution time(T$_{new}$) = $(\sum_{Q=1}^{p+n}(\frac{\sum_{i=1}^{f} T_{Qi}}{f}))/(p+n)$ |
| 25 | if avg(T$_{new}$) < avg(T$_{base}$) then A = P' |
| 27 | else Revert back and no change in P, A=P |
| 28 | Output *A* |

Figure 5. Knowledge Graph Adaptive Partitioning Algorithm.

The statistics use other feature patterns, such as SSJ, OOJ and OSJ and distributed joins in queries. The statistics comprise of (1) out degree sequence (hops) starting from the key feature (q) in a query graph pattern and its successive (peer) feature (p) present in the sequence, (2) triple size ratio (s) of the key feature and its successive (peer) features in shards and in the complete dataset, and (3) distributed joins in the queries. To balance the partition, the algorithm uses the statistics to determine the out degree of other features in the query to the key feature. It also uses features that are not involved in the workload, but present in the dataset. The algorithm monitors the query execution time and stores the statistics. It outputs the changes to shard compositions, based on the above information. Triples associated with the selected features are moved between shards and the partition metadata is updated. This operation is infrequent, and we assume that the system adjusts the partitioning only after identifying a significant change in the workload processing (the system monitors the execution time for each query). Typically, once the execution time increases significantly (given a threshold) the current partitioning is modified and an exchange of triples takes place. Queries from the new workload run according to the updated partition metadata and the runtime of the queries are being recorded.

## IV. IMPLEMENTATION

AWAPart stores an RDF dataset by partitioning it into sub-graphs, based on the initial query workload, and distributing the sub-graphs as shards among the nodes in a cluster. As the query workload changes, AWAPart establishes a new partitioning optimized for the new workload and dynamically adjusts the shards by triggering exchanges of subsets of triples between shards. The system is deployed on a single Master Node which controls the adaptive partitioning and a set of independent, share-nothing Processing Nodes, each with an installed triple store and a SPARQL query processor. The Master Node (Figure 6) is responsible for the overall workload analysis. It also controls the movement of triples subsets among the nodes in the cluster to adjust the partitioning. As the Master Node receives the query, the QueryAnalyzer and Feature Extractor (QAFE) starts the query feature extraction and updates the feature metadata. The Partition Manager (PM) uses the Hierarchical Agglomerative Clustering (HAC) module to cluster the extracted features. Using this HAC information, the Partition Metadata (PMeta) is updated. The dataset is indexed (IS) and according to PMeta, triples are searched and stored as shards. These shards are being uploaded to the processing nodes for the first time. A new query is sent to Query Rewriter and Processor (QRP), which rewrites the query into a federated query, based on the Partition Metadata (PMeta).

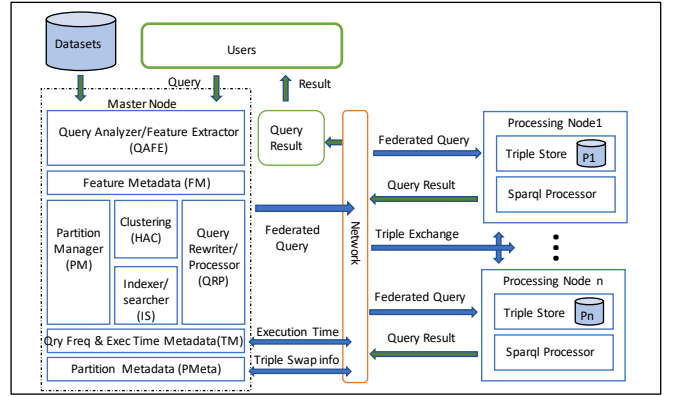

Figure 6. AWAPart System Architecture

This federated query is then sent to the processing node where it is going to be executed. The node where the query is executed is called the Primary Processing Node (PPN). The PPN is selected to minimize the distributed joins by selecting the shard with the highest number of features for the query. Adjustment of the partitioning of the RDF data is triggered by the Partition Manager (PM), due to changes in the workload query set and/or query frequency. The PM computes a new partition and, if the current shards require modifications, triples with selected features are exchanged between Processing Nodes to achieve a desired partitioning. The metadata of each Processing Node that was involved in triple swaps is updated to reflect the current state of triples in the shards. The PM uses the information stored in the Query Frequency and Execution Time Metadata (TM) and in the Feature Metadata (FM) with clustering information given by the Clustering Unit (HAC) to update Partition Metadata (PMeta). TM stores the information of every unique query and its average runtime.

## V. EXPERIMENTS

The synthetic dataset and queries in the Lehigh University Benchmark (LUBM [22]) were used for the evaluation of AWAPart, our knowledge graph adaptive partitioning method based on a query workload. LUBM includes basic information organized as a knowledge graph about a set of universities and related entities. It includes a set of 14 SPARQL queries intended for benchmarking of knowledge graph storage/query systems. The experiments were conducted on a cluster of Intel i5-based systems running Linux Ubuntu 18.04.4 LTS 64-bit OS. A relatively small cluster was selected to focus on the effects of repartitioning of the datasets of manageable sizes. There are many available RDF triple stores that provide the functionality of storing and querying the RDF data, such as Redland [23], Sesame, Jena [24], Virtuoso, etc. In the experiments, an instance of OpenLink Virtuoso [18] was installed on each node in the cluster. The knowledge graph partitioning and adaptive repartitioning systems, as well as the experiments were coded in Java with the use of the Apache Jena framework.

Two experiments were used to evaluate the effects of adaptive knowledge graph partitioning system, based on workload. (1) The first experiment was designed to evaluate the effectiveness of the adaptive partitioning to accommodate the changes in the set of queries in the workload. (2) The second experiment was created to evaluate the adaptive partitioning in response to the changes in the frequency of specific queries in the workload (the set of queries in the workload is unchanged, but some queries are executed more often than initially). An LUBM dataset of 10 universities, which included 1,563,927 triples was created and used. The initial partition [21] is created based on the initial query workload. The experiments show that the AWAPart system offers significant performance improvements over a system where the initial partitioning was unchanged.

**Experiment 1**: This experiment demonstrates the effects of changes in the composition of the workload query set on their performance, when executed on the initial partition and then on the adaptive partitioning. The changes to the workload included additions of new and/or deletions of existing queries. The modified workloads runtime on the initial partition and on our adaptive partitioning for the LUBM dataset were evaluated. Figure 7 shows 10 extra queries [25] EQ1 to EQ10 and 14 old queries Q1 to Q14 for the LUBM dataset and their runtimes. EQ1 to EQ10 are a mixture of linear, star, snowflake, and complex queries. The figures show the improvement in runtime performance for queries from EQ1 to EQ10 in milliseconds. Except for Q9, the performance of the other 13 original queries does not change. Figure 8 shows the average runtime of all 24 queries on the initial partition versus the adaptive partition in milliseconds. An overall improvement of 2 seconds of the adaptive partition over the

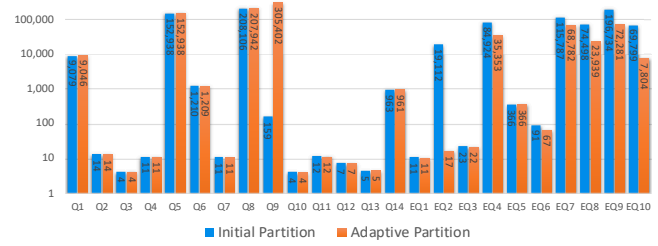
Figure 7. LUBM's 24 queries runtime in milliseconds

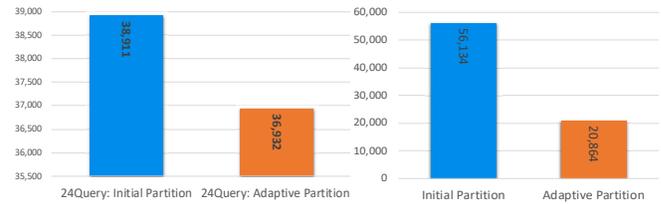
Figure 8. LUBM all 24-query average runtime in milliseconds

Figure 9. LUBM 10 new queries average runtime in milliseconds

initial partition is shown. Despite the drop in performance of a single query, the overall performance gains are clearly visible. If Q9 were replaced in the new workload composition (Figure 7), the performance gain would be even higher. Figure 9 shows that the improvement of the average runtime of the 10 new queries (EQ) on the initial partition is approximately 56 seconds, while the adaptive partition decreases it to 21 seconds. It is an improvement of 63% in the average runtime of the newly introduced queries on the adaptive partition over the initial partition. This experiment shows that the system can successfully adapt the partitioning with changes in the workload. At regular intervals, the system takes a snapshot of the current query workload and adapts the partitioning, which improves the workload runtime performance.

**Experiment 2**: This experiment examined the effects of the changes in the relative frequency of queries in the workload executed on the initial partition as compared to the adaptive partitioning and so the workload query frequency distribution

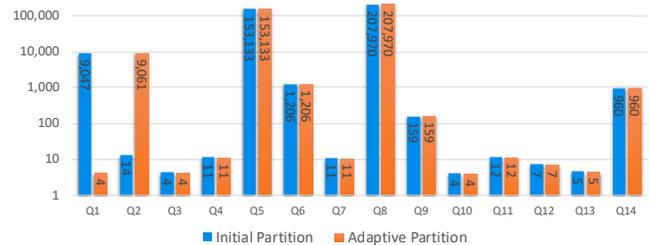
Figure 10. LUBM all queries average runtime of Initial vs. Adaptive partition in milliseconds

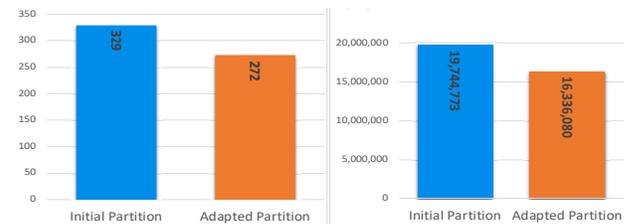
Figure 11. LUBM all query average runtime when frequency of Query1 is 50% of total workload a) Total runtime in minutes. b) Total runtime in milliseconds.

was altered. For example, if Q1 in LUBM is executed more frequently than the other 13 queries. The workload frequency share of query Q1 was increased to 50% of the whole workload. Figure 10 shows the changes in the runtime of queries Q1 and Q2. Queries 1 and 2 shares the same features. Our system swaps the queries based on score. This swapping reduces the distributed joins of Q1 but increases the distributed joins in the less frequently executed Q2, while maintaining the average runtime for the workload with evenly distributed queries. However, when the workload frequency is biased towards Q1, Figure 11 shows the improvement in the average workload performance by comparing the average runtime of the initial partition with biased workload frequency and adaptive partitioning with the biased workload frequency. The figure shows an improvement of approximately 17% of the adaptive partitioning over the initial partition, when the workload frequency is biased towards Q1.

The experiment shows that, when a query has a higher frequency than others, the performance of the adaptive partition against the initial partition is improved. Consequently, the system is adaptive to the changes in the workload. Again, at regular intervals, e.g., daily or after a set number of queries, the system takes a snapshot of the current query workload and query frequencies and, if needed, adapts the partitioning, which improves the average performance of the workload.

## VI. Conclusion And Future Work

In this paper, a system is proposed which is a distributed knowledge graph query processing system that adaptively partitions the graph according to changing workload. It aims to reduce the number of distributed joins during query execution that eventually leads to a reduced run-time for the queries achieving better performance. The system is adaptive with the new workload and the system learns the workload regularly and modifies the partition, which eventually improves the partition's overall runtime performance. Our experiments show the runtime comparison of workload aware initial partition versus adaptive partition. The results depict a significant increase in the performance of the queries. There is no need for replication of the data while optimizing the runtime of the workload queries.

In the future, a study of an evolving knowledge graph in terms of its schema and instances should be undertaken. Also, it will be interesting to examine how the adaptive partitioning handles the evolving datasets along with the evolving workload queries.